\documentclass[sigconf]{acmart}
\usepackage{xspace}
\usepackage{longtable}
\usepackage{array}
\usepackage[table]{xcolor}
\usepackage{graphicx}

\usepackage{color}
\usepackage{nameref}
\usepackage{multirow}
\usepackage{tabularx} 
\usepackage{censor}
\usepackage{enumitem}

\usepackage[most]{tcolorbox}

\usepackage{xcolor}          

\definecolor{lightblue}{RGB}{0, 0, 100}

\newtcolorbox{MyBox}{
  colback=white,
  colframe=lightblue,
  fonttitle=\bfseries,
  coltitle=black,
  sharp corners,
  boxrule=1pt,
  left=5pt,
  right=5pt,
  top=5pt,
  bottom=5pt,
  breakable
}

\definecolor{purplish}{HTML}{D8DFE3}
\definecolor{purplishlight}{HTML}{EBEFF3}
\definecolor{purplishdark}{HTML}{FF7F50}

\newtcolorbox[auto counter,number within=section]{rqbox}[2]{
    nameref=#1,
    title=\small{#1}, 
    enhanced,
    attach boxed title to top left={yshift=-6pt, xshift=8pt},
    boxed title style={size=small,boxsep=1pt},
    colframe=purplishdark,colback=white,colbacktitle=purplishdark,
    boxsep=2pt,left=2pt,right=2pt,top=6pt,bottom=2pt,middle=2pt
}

\AtBeginDocument{%
  }

\copyrightyear{2026}
\acmYear{2026}
\setcopyright{cc}
\setcctype{by}
\acmConference[CHASE'26]{19th International Conference on Cooperative and Human Aspects of Software Engineering}{April 12--18, 2026}{Rio de Janeiro, Brazil}
\acmBooktitle{19th International Conference on Cooperative and Human Aspects of Software Engineering (CHASE'26), April 12--18, 2026, Rio de Janeiro, Brazil}
\acmPrice{}
\acmDOI{10.1145/3794860.3794898}
\acmISBN{979-8-4007-2484-8/2026/04}

\begin{document}

\title{Empathy in Software Engineering Education: Evidence, Practices, and Opportunities}

\author{Matheus de Morais Leca}
\email{matheus.demoraisleca@ucalgary.ca}
\orcid{0009-0003-6596-9955}
\affiliation{%
  \institution{University of Calgary}
  \city{Calgary}
  \state{Alberta}
  \country{Canada}}

\author{Kim Johnston}
\email{johnstka@ucalgary.ca}
\orcid{0009-0009-7134-8876}
\affiliation{%
  \institution{University of Calgary}
  \city{Calgary}
  \state{Alberta}
  \country{Canada}}

\author{Ronnie de Souza Santos}
\email{ronnie.desouzasantos@ucalgary.ca}
\orcid{0000-0003-3235-6530}
\affiliation{%
  \institution{University of Calgary}
  \city{Calgary}
  \state{Alberta}
  \country{Canada}
  }

\begin{abstract}
\textbf{Context:} Empathy is increasingly recognized as a critical human capability for software engineers, supporting collaboration, ethical awareness, and user-centered design. While many disciplines have long explored empathy as part of professional formation, its incorporation into software engineering education remains fragmented. \textbf{Aim:} This study investigates how empathy has been used, taught, and discussed in general engineering and software engineering education, with the goal of identifying pedagogical practices, outcomes, and disciplinary differences that inform the structured integration of empathy into software curricula. \textbf{Method:} Following established guidelines for systematic reviews in software engineering, we conducted a comprehensive search across six databases and analyzed 43 primary studies published between 2001 and 2025. Data were coded and synthesized using descriptive and thematic analysis to capture how empathy is conceptualized, fostered, and assessed across educational contexts. \textbf{Findings:} Our findings show that engineering programs frame empathy as an ethical and reflective capacity linked to social responsibility, whereas software engineering translates empathy into structured, design-oriented, and measurable practices. Across both domains, empathy teaching enhances collaboration, ethical reasoning, bias awareness, and motivation, but remains limited by low curricular prioritization, measurement challenges, and resource constraints. \textbf{Conclusion:} Empathy is evolving from a peripheral soft skill into a measurable pedagogical construct in software engineering education. Embedding empathy as a continuous, assessable component of design and development courses can strengthen inclusivity, ethical reflection, and responsible innovation in future software professionals.
\end{abstract}


\received{20 October 2025}


\maketitle

\newcommand{\searchStrings}{3\xspace}
\newcommand{\topResults}{100\xspace}
\newcommand{\totalResults}{400\xspace}
\newcommand{\initialResults}{100\xspace}
\section{Introduction}
\label{sec:introduction}
Software engineering is not limited to technical competence alone. The discipline increasingly demands that professionals develop strong interpersonal abilities that support collaboration, communication, and ethical decision-making in complex environments \cite{matturro2019systematic, malinen2025soft, khakurel2020effect, lecca2025curious}. As software systems become more pervasive and socially embedded, professionals must balance analytical precision with human awareness \cite{albusays2021diversity}. Research consistently highlights that the success of software projects depends not only on technical expertise but also on the capacity of engineers to understand people, cooperate effectively, and respond to users’ diverse needs \cite{malinen2025soft, matturro2019systematic, khakurel2020effect}. These abilities, often referred to as soft skills, have been increasingly recognized as essential to software development and engineering education \cite{khakurel2020effect, lecca2025curious}.

Among these soft skills, empathy has emerged as an important human capability that supports teamwork, system design, and communication \cite{cerqueira2023thematic, cerqueiraexploring, gunatilake2023empathy, levy2018importance}. Empathy involves understanding others’ perspectives and emotions, fostering social connection and mutual respect \cite{gunatilake2023empathy, cerqueiraexploring}. Within software development, empathy contributes to clearer elicitation of requirements, improved collaboration, and a better understanding of the end-user's context \cite{ferreira2015eliciting, levy2018importance, devathasan2024deciphering}. Empathy has also been associated to healthier team dynamics and greater inclusion, as it supports psychological safety and mitigates conflict in diverse environments \cite{devathasan2025empathy, cerqueiraexploring}. Despite this growing recognition, empathy remains underexplored in software engineering curricula, where technical proficiency often takes precedence over interpersonal development \cite{cerqueiraexploring, cerqueira2023thematic}.

Empathy has gained increasing attention in engineering education as an essential component of professional formation, closely linked to human-centered design, teamwork, and social responsibility among students \cite{hess2012empathy, khakurel2020effect, levy2018importance}. Despite this recognition, the incorporation of empathy into software engineering education remains limited and fragmented, with studies offering diverse yet disconnected perspectives \cite{cerqueiraexploring, cerqueira2023thematic, gunatilake2023empathy}. The absence of synthesis across these efforts hinders the systematic integration of empathy as a structured element of software engineering curricula. To address this gap, this study investigates how empathy has been incorporated in both engineering education, in general, and software engineering education, in particular, aiming to identify practices, reported benefits, challenges, and opportunities for adaptation across domains. Accordingly, the research is guided by the following question: \textbf{RQ: \textit{How has empathy been used, taught, and discussed in engineering and software engineering education?}} Overall, the contributions of this paper are:
\begin{enumerate}
\item \textbf{Mapping Study}. A synthesis of 43 studies addressing empathy in engineering (in general) and software engineering education, identifying how empathy has been used, taught, and evaluated across different contexts, including reported benefits and challenges.

\item \textbf{Comparative Analysis}. A cross-disciplinary comparison showing that engineering education frames empathy as an ethical and reflective capacity, while software engineering emphasizes structured, design-oriented, and assessable practices that can inform curriculum design.

\item \textbf{Educational Baseline}. A set of evidence-based practices that provide a foundation for systematically integrating empathy into software engineering education.
\end{enumerate}

From this introduction, this paper is organized as follows. Section~\ref{sec:method} describes our methodological approach. Section~\ref{sec:findings} presents the synthesis across four dimensions: use of empathy, benefits, challenges, and methods to foster empathy. Section~\ref{sec:discussion} discusses implications for educational practice. Finally, Section~\ref{sec:conclusions} summarizes the conclusions and proposes directions for future work.

\section{Theoretical Background} \label{sec:background}

This section presents two core concepts in this study: empathy in software engineering practice and empathy as a pedagogical foundation for professional formation.

\subsection{Empathy as a Software Engineering Skill}

Empathy has been increasingly recognized as a necessary human capacity for addressing the social dimensions of software engineering practice. Studies focusing on this theme describe empathy as a professional attribute that enables engineers to connect with the perspectives, emotions, and needs of the users \cite{hess2012empathy, khakurel2020effect, cerqueira2023thematic}. Within this view, empathy is not only a moral disposition but also a cognitive and behavioral process that supports understanding user contexts, enhancing teamwork, and promoting ethical decision-making. This emphasis reinforces the view of software engineering as a discipline that integrates human awareness into design, communication, and decision processes rather than focusing solely on technical aspects of software \cite{malinen2025soft, matturro2019systematic, lecca2025curious}.

In the broader software engineering literature, empathy is often positioned as both a soft skill and a professional competency that can be cultivated through intentional practice \cite{cerqueiraexploring, cerqueira2023thematic, cerqueira2024empathy, clear2024software, devathasan2025empathy}. Research on soft skills highlights that empathy, together with communication and collaboration, is critical for professional success and project outcomes \cite{malinen2025soft, matturro2019systematic, lecca2025curious, khakurel2020effect, gunatilake2023empathy}. These skills complement analytical reasoning by allowing software engineers to map complex human and organizational interactions into code \cite{ferreira2015eliciting, levy2018importance, cerqueiraexploring, gunatilake2023empathy}. However, despite being widely acknowledged, empathy is rarely systematically assessed and remain ambiguous.

Within software engineering, the concept of empathy appears prominently in human-centered and design thinking approaches. Design thinking emphasizes understanding user experiences as a foundation for innovation, making empathy a key step in framing problems and identifying user needs \cite{levy2018importance, cerqueira2023thematic}. In software contexts, techniques such as empathy maps, personas, and user scenarios translate emotional and contextual insights into design requirements, helping bridge the gap between developers and users \cite{ferreira2015eliciting, cerqueiraexploring, gunatilake2023empathy}. These practices encourage software engineers to approach requirements elicitation as an interpretive and human-centered process rather than a purely analytical one, aligning with contemporary views that effective software design emerges through iterative engagement with diverse stakeholders and continuous reflection on user perspectives \cite{cerqueira2023thematic, clear2024software, gunatilake2024enablers}.

Empathy within professional software practice also extends beyond individual interactions to collective and organizational levels. Empirical studies show that empathy enables developers to communicate more effectively, understand user challenges, and balance technical priorities with ethical considerations \cite{gunatilake2023empathy, gunatilake2024enablers, clear2024software}. Lack of empathy, in contrast, contributes to user alienation, overlooked privacy concerns, and inequitable design choices \cite{levy2018importance, cerqueira2023thematic}. Research also suggests that empathy within teams, sometimes described as collective empathy, plays an important role in fostering inclusion, reducing conflict, and supporting motivation in diverse environments \cite{devathasan2024deciphering, devathasan2025empathy}. These findings position empathy as both a social and cognitive mechanism that strengthens ethical responsibility and cohesion in collaborative software development.

\subsection{Teaching Empathy and Pedagogical Empathy}
Empathy has long been regarded as a learnable and essential component of professional education \cite{meek1957experiment, thompson1983empathy, stepien2006educating, jeffrey2016empathy}. Across decades of research, the motivation to teach empathy arises from the recognition that human connection supports effective communication, ethical awareness, and professional competence \cite{meek1957experiment, thompson1983empathy, stepien2006educating, jeffrey2016empathy}. Educational programs in many fields have sought to develop empathy as both an interpersonal skill and a moral disposition that allows professionals to understand the emotions and perspectives of others \cite{han2018review, kelley2013teaching}. Overall, the broader goal of empathy education has been to cultivate practitioners who cannot only perform technical duties but also respond to the human experiences that surround their professional actions \cite{bearman2015learning, van2012based}. 

Teaching empathy has been particularly developed in healthcare education, where research demonstrates that empathic practice improves patient outcomes, satisfaction, and trust \cite{han2018review, stepien2006educating, jeffrey2016empathy}. Medical educators have emphasized that empathy can be taught through structured activities that engage emotional, cognitive, and behavioral dimensions, such as communication workshops, simulations, narrative reflection, and role-playing exercises \cite{stepien2006educating, jeffrey2016empathy}. Studies show that these practices not only improve professional–patient interactions but also help to counter the well-documented decline in empathy during medical training \cite{han2018review, jeffrey2016empathy}. Nursing education similarly emphasizes experiential learning, active listening, and compassion-based communication, demonstrating that empathy training strengthens both professional effectiveness and emotional resilience \cite{kelley2013teaching}. 

Historical and contemporary educational research shows that empathy has been approached as both a pedagogical aim and a transferable skill \cite{meek1957experiment, thompson1983empathy}. Early classroom interventions demonstrated measurable gains in empathic ability when teaching included reflective dialogue, group discussion, and peer observation \cite{meek1957experiment}. Later, empathy training was expanded to humanities-based disciplines, where creative activities such as storytelling, dramatization, and historical perspective-taking were used to cultivate understanding of diverse human experiences \cite{thompson1983empathy}. These approaches highlight empathy not merely as an emotional reaction but as an intellectual and imaginative engagement that supports ethical awareness and social understanding within education.

Recent developments extend empathy education beyond healthcare and social studies into professional and technical programs. Studies in professional education demonstrate that design-oriented exercises can foster empathy through narrative and user-centered approaches \cite{van2012based}. For example, the construction of research-based personas enables students to imagine the needs, emotions, and motivations of others, strengthening their capacity for perspective-taking and ethical reflection in professional contexts. Similarly, collaborative and experiential learning models emphasize curiosity, reflection, and mindfulness as strategies for developing empathy that endures beyond the classroom \cite{jeffrey2016empathy, bearman2015learning}. These findings indicate that empathy can be cultivated through intentional design of learning experiences that connect knowledge with human values.

Globally, calls to teach empathy reflect a response to growing social fragmentation, misinformation, and polarization. Educational institutions are increasingly viewed as critical spaces for fostering mutual understanding, social responsibility, and ethical citizenship \cite{gates2023world, jeffrey2016empathy, han2018review}. Across disciplines, scholars emphasize that empathy education can counter detachment and promote inclusive and reflective learning environments, helping learners appreciate different perspectives and act with moral sensitivity \cite{bearman2015learning, stepien2006educating, thompson1983empathy}. Teaching empathy prepares students to engage constructively across cultural and disciplinary boundaries, equipping them with the capacity to navigate complex interpersonal and societal challenges \cite{kelley2013teaching, van2012based}. In this sense, pedagogical empathy is not confined to any single domain but represents a foundational principle for education that seeks to humanize learning and professional practice \cite{meek1957experiment, davis1990empathy, jeffrey2016empathy}.
\section{Method}
\label{sec:method}
This study follows established methodological principles for systematic reviews, drawing on widely recognized guidelines in software engineering research \cite{kitchenham2009systematic} and aligned with contemporary empirical standards for evidence-based studies \cite{ralph2020empirical}. The research initially focused exclusively on software engineering education, aiming to understand how empathy was being incorporated and taught within this field. However, the initial search revealed a limited number of studies directly addressing empathy in software engineering contexts. As we expanded our exploration, it became evident that other branches of engineering had engaged more actively with empathy-related research. Consequently, the scope of our review was broadened to include engineering education more generally, allowing for comparative analysis and the identification of practices, concepts, and pedagogical approaches that remain underexplored in software engineering. The overarching goal of the review, then, focused on conducting a structured synthesis that investigates how empathy is incorporated across engineering disciplines and within software engineering in particular.

\subsection{Research Questions}
Following our overarching goal and general research question, we devised four specific research questions to guide this review:

\begin{enumerate}
\item \textbf{RQ1:} How is empathy being incorporated in engineering education in general and in software engineering education in particular?
\item \textbf{RQ2:} What methods are being used to foster empathy in engineering education in general and in software engineering education in particular?
\item \textbf{RQ3:} What are the reported benefits of applying empathy in engineering education in general and in software engineering education in particular?
\item \textbf{RQ4:} What are the challenges of applying empathy in engineering education in general and in software engineering education in particular?
\end{enumerate}

These questions were developed to align with the study’s objective to support a systematic synthesis of pedagogical practices, observed benefits, and identified challenges, offering a structured foundation for understanding how empathy can be effectively integrated into software engineering education.

\subsection{Search Strategy}
Following the guidelines of systematic literature reviews \cite{kitchenham2009systematic, ralph2020empirical}, we adopted a five-step protocol: (i) planning, (ii) search, (iii) selection, (iv) data extraction, and (v) synthesis. The search strategy combined automated and manual techniques to ensure comprehensive coverage and to mitigate potential sampling and publication bias. Searches were conducted across six major electronic databases frequently used in software engineering and education research: \textit{Google Scholar}, \textit{IEEE Xplore}, \textit{ACM Digital Library}, \textit{Compendex}, \textit{Science Direct}, and \textit{Scopus}. The search string was defined iteratively during pilot searches and finalized as follows:

\begin{quote} \texttt{("empathy" OR "empathetic" OR "perspective taking" OR "compassion") AND ("engineering education" OR "engineering curriculum" OR "engineering pedagogy" OR "engineering teaching" OR "engineering learning")}
\end{quote}

\subsection{Inclusion and Exclusion Criteria}
To ensure the selection of relevant and high-quality studies, we applied the following criteria:

\textbf{Inclusion Criterion (IC):}
\begin{itemize}
    \item \textbf{IC1:} Focus on studies examining the application of empathy concepts within the learning processes of engineering programs, explicitly discussing enablers, benefits, or limitations.
\end{itemize}

\textbf{Exclusion Criteria (EC):}
\begin{itemize}
    \item \textbf{EC1:} Do not discuss the application of empathy concepts in the learning processes of engineering programs or do not have fostering/teaching empathy as the main focus.  
    \item \textbf{EC2:} Are not full papers (e.g., dissertations, theses, monographs, short papers under five pages).  
    \item \textbf{EC3:} Cannot be accessed through institutional credentials.  
    \item \textbf{EC4:} Are not written in English.  
    \item \textbf{EC5:} Are duplicate studies.
\end{itemize}

Each study was independently screened by two reviewers, with discrepancies resolved through discussion to maintain rigor and reliability in the selection process.

\subsection{Data Collection and Extraction}
Our search process initially identified 374 studies. After removing 38 duplicates (EC5), 336 unique studies remained for title and abstract screening. From these, 220 studies were excluded for not discussing empathy in engineering education (EC1), leaving 116 for full-text review. At the full-text stage, 33 were excluded for being non-full papers (EC2), 21 for inaccessibility (EC3), and 1 for language (EC4). Consequently, 43 studies met IC1 and were included in the final synthesis. These papers are listed in Table \ref{tab:included-studies}. For each included study, we extracted metadata such as authors, year, publication venue, and the specific engineering discipline represented. We also recorded analytical variables directly related to the research questions, including: (i) how empathy was defined or conceptualized, (ii) the educational strategies or methods used to foster it, (iii) the reported benefits of its application, and (iv) the challenges or barriers identified. The extraction process was conducted using a structured form designed to ensure uniformity and traceability across all reviewed studies. Each paper was read in its entirety, and key details were summarized to create a comprehensive and transparent dataset that supported subsequent synthesis and comparison across disciplines.

\begin{table*}[t]
\centering
\caption{List of studies included in the review}
\label{tab:included-studies}
\footnotesize
\begin{tabular}{p{0.7cm} p{15cm}}
\toprule
\textbf{ID} & \textbf{Title} \\
\midrule
A01 & Role of Empathy in Engineering Education and Practice in North America \\
A02 & Caring for the Future: Empathy in Engineering Education to Empower Learning \\
A03 & Reframing Engineering Design: Integrating Design Thinking and Systems Thinking in Engineering Education and Practice to Address Wicked Problems \\
A04 & Engineering Students’ Empathy Development through Service Learning: Examining Individual Student Experiences in a Technical Communication Course \\
A05 & Efficacy of Humanities-Driven Science, Technology, Engineering, and Mathematics Curriculum on Integrating Empathy into Technology Design \\
A06 & Educating for Empathy in Software Engineering Course \\
A07 & Empathy-Driven Student Transformations: Bridging the Gap in Software Development for Inclusive User Experiences \\
A08 & Video-Based Empathy Training for Software Engineers \\
A09 & Model and Methodology for Developing Empathy: An Experience in Computer Science Engineering \\
A10 & Open Questions for Empathy-Building Interventions for Inclusive Software Development \\
A11 & Personas and Scenarios as a Method to Develop Empathy and Teamwork in Computer Engineering Education \\
A12 & Board 175: Poster: Strategies for Empathy Instruction and Assessment in Biomedical Engineering Education: A Review \\
A13 & Transdisciplinary STEAM Education: Advocating for Compassion as a Core Value in Engineering \\
A14 & Review on Empathy in Engineering Education: Conceptions, Interventions, and Challenges \\
A15 & Needed: A Few Good Knights for the Information Age – Competence, Courage, and Compassion in the Engineering Curriculum \\
A16 & Psychological Safety and Empathy in Collaborative Learning Environments: A DEI Strategy for Engineering Education \\
A17 & From ‘Empathic Design’ to ‘Empathic Engineering’: Toward a Genealogy of Empathy in Engineering Education \\
A18 & Fostering Empathy in an Undergraduate Mechanical Engineering Course \\
A19 & Engineering Students’ Empathy Development through Service Learning: Quantitative Results from a Technical Communication Course \\
A20 & Emotional Tool Design: A Perspective on How to Generate Empathy in Higher Education Students \\
A21 & A Quantitative, Pilot Investigation of a Service-Learning Trip as a Platform for Growth of Empathy \\
A22 & Integrating Large Language Models into Engineering Education: Redesigning Peer Feedback Assessments for Enhanced Learning Outcomes \\
A23 & WIP: Contemplative Practices’ Effects on Compassion, Belonging, and Self-Empowerment in Undergraduate Engineering Experiences \\
A24 & Contemplating Engineering and Science: Creating Compassionate and Empathetic Learning Spaces in Engineering Education \\
A25 & Empathy Enhancement among Engineering Students through Cooperative Problem-Based Learning \\
A26 & Role of Empathy to Identify Unmet Needs in Indian Villages through Service Learning Program – A Case of the Unnat Bharat Abhiyan Program \\
A27 & Effects of Design Thinking on Transnational Collaborative Projects in Engineering \\
A28 & Virtual Reality to Enhance User Satisfaction in an Engineering Innovation Project \\
A29 & The People Part of Engineering: Engineering for, with, and as People \\
A30 & Emotional Intelligence in Engineering Education: Exploring the Influence of Empathetic Design Approaches in a Fourth-Year Engineering Class \\
A31 & Developing Novel Practices of Somatic Learning to Enhance Empathic Perspective-Taking for Ethical Reasoning and Engineering Design \\
A32 & Reflective Practitioners through Design: Perspectives of Second-Year Engineering Undergraduate Students \\
A33 & Engineering Students’ Empathy Development through Service Learning: Qualitative Results in a Technical Communication Course \\
A34 & Empathy as Key to Inclusivity in Engineering Education \\
A35 & The Impact of Student-Centered Learning towards Reinforcement of Positive Values among Chemical Engineering Students \\
A36 & On Positionality and the Implementation of Experiential Learning for Engineers \\
A37 & Using Accessibility Awareness Interventions to Improve Computing Education \\
A38 & Where We Are: Understanding Instructor Perceptions of Empathy in Engineering Education \\
A39 & Work in Progress: Fostering Empathetic Engineers by Practicing Contextual Listening – A Case Study \\
A40 & Empathy and Ethical Becoming in Biomedical Engineering Education: A Mixed Methods Study of an Animal Tissue Harvesting Laboratory \\
A41 & Presenting and Evaluating the Impact of Experiential Learning in Computing Accessibility Education \\
A42 & Enhancing Empathy and Innovation in Engineering Education through Design Thinking and Design of Experiments \\
A43 & Empathy, Persuasiveness and Knowledge Promote Innovative Engineering and Entrepreneurial Skills \\
\bottomrule
\end{tabular}
\end{table*}

\subsection{Quality Analysis}

To ensure that the synthesis was based on reliable and methodologically sound evidence, we conducted a quality analysis of all papers included after the screening phase. The assessment focused on two dimensions: (i) whether each study employed methods grounded in well-established research guidelines and literature, and (ii) whether the results reported were consistent with the described methods. Each study was rated on a binary scale: papers that satisfied both conditions were marked as having sufficient quality for inclusion in the mapping; those that did not were excluded. The assessment was independently performed by two reviewers. Discrepancies were discussed and resolved by consensus. After this process, all 43 papers were confirmed as meeting the defined quality criteria and were included in the final synthesis.

\subsection{Data Analysis and Synthesis}
After the extraction stage, we applied two complementary approaches to synthesize the findings. First, descriptive analysis provided an overview of research trends \cite{george2018descriptive}, while thematic synthesis \cite{cruzes2011recommended} offered an interpretive understanding of how empathy has been incorporated into teaching and learning within engineering and software engineering education. \textbf{Descriptive Statistics.}
The first stage involved a quantitative summary of the selected studies. The analysis followed general procedures for descriptive statistical analysis in empirical research \cite{george2018descriptive}. Data were organized by year, discipline, and research focus to characterize the evolution of the field. Frequency counts and percentages were calculated to capture distribution patterns and proportions. These measures were presented in tables and visual summaries to highlight disciplinary representation and temporal development. The resulting descriptive overview provided a quantitative foundation for identifying the growth and concentration of research on empathy in engineering and software engineering education. \textbf{Thematic Synthesis.}
The qualitative synthesis followed established procedures for thematic analysis in software engineering research, as described in the methodological literature on evidence-based synthesis \cite{cruzes2011recommended}. The process involved three main phases:

\begin{enumerate}
\item \textbf{Identifying Concepts:}
Qualitative coding techniques were applied to extract and label text segments describing how empathy was incorporated, taught, or assessed in each study. This phase resembled open coding, allowing concepts to emerge inductively from the data. Each study was read in full, and excerpts related to the research questions were coded to capture information about empathy-related practices, outcomes, and challenges. Similar codes were compared and aligned across studies to ensure consistency in meaning.

\item \textbf{Grouping Concepts:}  
The second phase resembled axial coding and involved organizing related codes into broader descriptive categories. The grouped codes formed themes representing recurring ideas across studies, such as methods for teaching empathy, perceived learning benefits, and barriers to implementation. These descriptive themes were refined through repeated comparison and review to ensure coherence and accurate representation of the data.  

\item \textbf{Developing Analytical Themes:}  
The final phase followed a process similar to selective coding, in which relationships among the descriptive themes were analyzed to develop higher-order analytical themes. These themes explained how empathy has evolved from a moral and reflective construct to a measurable and teachable competence in engineering and software engineering education, which is described in our discussions.  

\end{enumerate}

The integration of descriptive and thematic analyses provided both scope and depth. Descriptive statistics offered a structured overview of research activity across disciplines, while thematic synthesis enabled interpretation of the pedagogical and conceptual patterns identified.

\subsection{Threats to Validity}
Systematic reviews are subject to several validity threats that can affect the reliability and interpretation of their findings. In this study, potential sources of bias were considered throughout the process, from study selection to data synthesis. A possible limitation concerns database coverage and publication language. To reduce this risk, the search strategy combined multiple databases that cover education and software engineering research. Cross-verification between search results and reference lists helped ensure that relevant studies were not omitted. Screening and eligibility assessments were conducted independently by two authors, and any disagreements were resolved through discussion in conflict meetings until consensus was reached.

Researcher interpretation can also influence the coding and synthesis of qualitative data. To address this, the review followed a structured extraction form and a transparent coding process aligned with established practices for thematic synthesis in software engineering research \cite{cruzes2011recommended}. The two authors collaboratively developed the coding framework and refined it through iterative discussions, merging overlapping categories and verifying consistency in code application. The combination of inductive and deductive reasoning supported both exploratory insights and alignment with the research questions.

Finally, to promote transparency and reproducibility, the full review protocol, inclusion criteria, and coding framework are provided in the paper and as supplementary material. These resources allow replication and evolution of our review. The integration of descriptive and thematic analyses offered methodological triangulation between quantitative and qualitative findings, strengthening the validity and credibility of the conclusions, while recognizing that variations in study reporting may still limit the completeness of the available evidence.
\section{Results} 
\label{sec:findings}

Our mapping identified research on empathy in engineering education in general, spanning 24 years, from 2001 to 2025. However, this period was not continuous, as the first paper appeared in 2001, while the next publications did not emerge until 2014, resulting in a 13-year gap with no studies conducted between 2002 and 2013. Across the entire period, 43 papers were identified across six engineering domains. The areas represented include Multidisciplinary, Software, Civil, Chemical, Biomedical, Mechanical, and Computer Engineering. The multidisciplinary group dominates the field with 23 papers, and five of them explicitly integrate systems perspectives. Other disciplinary representations are smaller: Software (9 papers), Civil (3), Chemical (3), Biomedical (3), Mechanical (1), and Computer (1). Table \ref{tab:area-summary} presents the temporal distribution of papers across areas and years, showing that research on empathy in engineering began as a broad pedagogical effort and later expanded into more specialized domains.

Nowadays, within these domains, software engineering education has given the most attention to empathy. Research on educational empathy in this area emerged much later, with the first study published in 2021, approximately 20 years after the initial work in engineering education. Before that, other fields had already introduced empathy in their teaching research, including Multidisciplinary Engineering studies in 2014, Chemical Engineering in 2015, Mechanical Engineering in 2016, and Biomedical Engineering in 2017. Between 2021 and 2025, nine studies were identified in Software Engineering and one in Computer Engineering. Thus, a total of ten studies related to software, which represent about 23 percent of all papers in the mapping.

A comparison of these numbers highlights a clear shift. Empathy in engineering education has been introduced through general and design-focused initiatives, but software-related disciplines have recently become central to the conversation. Engineering education as a whole took over a decade to establish continuity in the research on empathy, while software education reached similar publication levels in less than five years. This rapid growth reflects the increasing recognition of empathy as an important competence for responsible and inclusive technology development. Based on this evolution, the following sections address our research questions in detail. While the papers from which the evidence was drawn are mapped to the description of our findings using their codes, not all studies appear in every section. During data collection, some papers did not provide explicit or implicit information about all aspects explored, such as uses, practices, benefits, or challenges. Therefore, their presence varies according to the relevance of the evidence identified.

\begin{table}[t]
\centering
\caption{Evolution of studies on empathy in Engineering Education (2001--2025)}
\label{tab:area-summary}
\footnotesize
\begin{tabular}{p{2cm} p{0.9cm} p{2cm} p{2cm}}
\toprule
\textbf{Area} & \textbf{Count} & \textbf{Years Covered} & \textbf{IDs} \\
\midrule
Multidisciplinary & 18 & 2001, 2016--2025 & A04, A10, A13, A14, A15, A16, A17, A18, A20, A21, A22, A23, A26, A30, A32, A36, A37, A39 \\
\midrule
Multidisciplinary (includes systems) & 5 & 2014--2024 & A06, A25, A28, A29, A31 \\
\midrule
Software & 9 & 2021--2024 & A02, A03, A05, A09, A34, A38, A40, A42, A43 \\
\midrule
Civil & 3 & 2020--2022 & A01, A24, A33 \\
\midrule
Chemical & 3 & 2015--2023 & A11, A27, A41 \\
\midrule
Biomedical & 3 & 2017--2025 & A08, A12, A35 \\
\midrule
Mechanical & 1 & 2019 & A19 \\
\midrule
Computer & 1 & 2025 & A07 \\
\bottomrule
\end{tabular}
\end{table}

\subsection{RQ1: Incorporating Empathy in Engineering and Software Engineering Education}

In engineering education, empathy appears as an ethical and reflective professional competency that supports social responsibility and human-centered awareness (A08, A10, A13, A14, A15). Integration occurs through conceptual and cultural approaches such as reflection, ethics discussions, and professional identity formation (A13, A18, A31). The goal remains to develop socially aware engineers capable of understanding the human and societal implications of their work (A01, A32, A37). Across studies, empathy evolves from a moral value associated with personal traits to a trainable professional skill (A13, A15, A18, A19).

In software engineering education, empathy is incorporated as a design-oriented and measurable practice embedded in problem definition, user research, and system design (A02, A03, A34, A38). Application occurs through empathy maps, stakeholder interviews, and inclusive design activities, promoting perspective-taking toward diverse users, including those with disabilities or attention disorders (A02, A03, A34). Within this context, empathy functions as both a collaborative competence that supports teamwork and communication (A05, A09, A38) and a tool for ethical and accessible software creation (A02, A05, A34, A40). The evidence shows a progression from a supporting design aid to a foundational element of user-centered and ethical software development (A02, A03, A05, A34, A38).

Both groups of evidence reveal a shared movement toward legitimizing empathy as a central element of engineering education. However, while engineering in general emphasizes empathy’s ethical and professional dimensions, software engineering translates those values into practical design actions and measurable learning outcomes. Empathy therefore operates at two complementary levels: as a moral compass shaping engineers’ sense of responsibility, and as a design instrument shaping the technologies software engineers create, as summarized in Table \ref{tab:comparison}.

\begin{table}[t]
\centering
\caption{Comparison of empathy incorporation in Engineering and Software Engineering Education}
\label{tab:comparison}
\footnotesize
\begin{tabular}{p{2cm} p{2.6cm} p{2.6cm}}
\toprule
\textbf{Aspect} & \textbf{Engineering (General)} & \textbf{Software Engineering} \\
\midrule
Nature of Empathy & Ethical, reflective, professional & Cognitive, affective, and design-oriented \\
\midrule
Pedagogical Focus & Human-centered ethics and social responsibility & User-centered design and inclusive technology \\
\midrule
Integration & Conceptual and cultural & Practical and method-driven \\
\midrule
Outcome & Develop socially aware engineers & Build software that reflects empathy \\
\bottomrule
\end{tabular}
\end{table}

\subsection{RQ2: Pedagogical Practices to Incorporate Empathy in Engineering and Software Engineering Education}

Across engineering education, diverse pedagogical practices are used to foster empathy in both classroom and experiential learning contexts. The most recurrent approaches involve reflective and ethical learning, where students engage in reflection assignments, ethics discussions, and critical reasoning exercises to connect technical problem-solving with moral and social awareness (A08, A11, A12, A13, A22, A26, A35, A37). Design thinking and empathic design approaches also play a central role, encouraging perspective-taking through iterative, user-centered problem framing and solution generation (A24, A28, A29, A30, A31, A32, A33, A35, A39).

Empathy is further promoted through service learning and real-user engagement, where community-based projects and humanitarian engineering activities help students experience social contexts firsthand (A11, A14, A15, A16, A20, A22, A25, A36). Experiential and simulation-based learning, including workshops, functional simulations, and laboratory activities, exposes students to real or simulated user challenges (A08, A10, A12, A15, A17, A21, A23, A33, A41). Collaboration is another recurring element, reflected in cooperative, problem-based, and interdisciplinary teamwork that builds mutual understanding across disciplinary and cultural boundaries (A11, A14, A15, A17, A18, A19, A27, A30, A33, A39).

Several studies propose framework-based and structured models for empathy education, outlining theoretical or curricular structures to guide integration across courses (A04, A08, A13, A30, A35). Others focus on assessment and measurement, employing instruments such as surveys, questionnaires, and mixed-methods evaluations to track empathy development (A06, A10, A20, A21). Finally, a number of efforts address educator preparation and support, emphasizing faculty empowerment and curriculum reform as mechanisms to sustain empathy-focused education (A01, A19, A24, A33).

In software engineering education, the same pedagogical foundations appear but with stronger emphasis on structured, measurable, and design-integrated activities. Reflective learning occurs within projects and teamwork through guided writing, feedback sessions, and multimodal reflection exercises (A02, A03, A05, A09, A34, A38). Design thinking is tightly embedded in requirements engineering, supported by empathy maps, personas, and stakeholder analysis to connect user perspectives to system functionality (A02, A34, A38). Service and real-user engagement also feature prominently, often through accessibility projects and client-based collaborations that expose students to authentic user experiences (A03, A34).

Software engineering introduces distinct forms of experiential and simulation-based learning, such as empathy labs, impairment simulations, and accessibility redesign tasks (A03, A34, A40, A42). Collaborative practice takes the form of team-based programming, peer empathy evaluation, and multidisciplinary work with design students (A02, A09, A38). Structured pedagogical frameworks, including the SPC methodology and modular empathy-building interventions, sequence empathy learning within software development cycles (A09, A34). Assessment practices are quantitative and embedded, relying on pre–post testing, artifact evaluation, and vocabulary analysis to monitor empathy progression (A05, A09, A34).

Both areas reveal increasing pedagogical sophistication in fostering empathy, yet they diverge in focus. Engineering education emphasizes broad, humanistic, and ethical approaches, whereas software engineering operationalizes empathy through structured, iterative, and data-informed practices aligned with technical development. These patterns highlight how empathy teaching evolved from conceptual exercises to discipline-embedded interventions, as summarized in Table \ref{tab:pedagogical-comparison}.

\begin{table}[t]
\centering
\caption{Comparison of pedagogical practices to incorporate empathy in Engineering and Software Engineering Education}
\label{tab:pedagogical-comparison}
\footnotesize
\begin{tabular}{p{2cm} p{2.7cm} p{2.7cm}}
\toprule
\textbf{Pedagogical Theme} & \textbf{Engineering (General)} & \textbf{Software Engineering} \\
\midrule
Reflective and Ethical Learning & Reflection and ethics-based activities encouraging moral reasoning and human-centered awareness & Guided reflections and feedback integrated into projects and teamwork \\
\midrule
Design Thinking and Empathic Design & Design thinking and empathic design methods emphasizing user perspective and contextual understanding & Design thinking embedded in software design through empathy maps, personas, and stakeholder analysis \\
\midrule
Service Learning and Real-User Engagement & Community and humanitarian projects connecting students with real stakeholders & Client-based experiential learning and accessibility-focused user collaboration \\
\midrule
Experiential and Simulation-Based Learning & Workshops, simulations, and laboratory experiences offering hands-on engagement & Empathy labs, impairment simulations, and accessibility redesign activities \\
\midrule
Collaborative and Team-Based Practice & Cooperative, problem-based, and interdisciplinary teamwork fostering peer empathy & Team-based programming, peer empathy evaluation, and multidisciplinary collaboration \\
\midrule
Framework-Based and Structured Models & Conceptual and curricular frameworks guiding empathy integration across courses & Structured frameworks sequencing empathy learning through defined stages and activities \\
\midrule
Assessment and Measurement & Surveys and mixed-methods evaluations assessing empathy development & Embedded pre/post testing, artifact analysis, and quantitative tracking of empathy progression \\
\midrule
Educator Preparation and Support & Faculty training and curriculum reform promoting empathy-focused instruction & Instructor feedback and expert project review reinforcing empathic design practices \\
\bottomrule
\end{tabular}
\end{table}

\subsection{RQ3: Benefits of Incorporating Empathy in Engineering and Software Engineering Education}

In engineering education, the incorporation of empathy in teaching has been shown to strengthen both learning and professional formation. The most consistent benefit across studies involves improved analytical and creative problem solving, as empathic approaches help students approach technical challenges with broader contextual awareness and social responsibility (A01, A04, A08, A13, A17, A20, A22, A25, A28, A29, A31, A33, A36, A39, A41). Teaching practices that include reflection, design thinking, and service learning help students develop more human-centered solutions and recognize engineering problems as socially situated (A08, A13, A17, A25, A33, A39).

Collaborative learning environments show that empathy teaching enhances teamwork and interpersonal understanding. Group projects, interdisciplinary work, and cooperative learning encourage communication and recognition of diverse perspectives and emotions (A01, A08, A17, A19, A27, A37, A39). Courses that link empathy to ethical reflection or community engagement report higher levels of professional and social awareness, helping students connect classroom learning with civic responsibility and the societal implications of engineering work (A01, A24, A32, A35, A37).

Teaching empathy also contributes to student motivation and engagement by linking technical content with socially meaningful learning experiences (A08, A16, A17, A29, A36, A37). Students who participate in empathy-based activities report a stronger interest in their studies and a greater identification with the social relevance of engineering practice. Additional benefits include increased self-awareness and communication skills valued in professional settings (A06, A10, A26, A37). Empathy-focused teaching also helps students recognize and mitigate bias, supporting inclusive and socially conscious design practices (A32, A33, A37). These educational strategies collectively help future engineers understand their broader responsibilities and the impact of their work in society (A16, A24, A25, A32, A37).

In software engineering education, empathy is incorporated into teaching as a foundation for user-centered and ethically responsible design. Pedagogical approaches such as user research, persona development, empathy mapping, and accessibility-focused projects enable students to understand real users’ needs and experiences (A02, A03, A34, A38). Through these practices, empathy improves students’ ability to design inclusive, accessible, and socially meaningful software systems. Collaborative activities such as multidisciplinary teamwork and peer review promote communication and shared understanding in design and programming teams (A02, A03, A05, A09, A38).

Empathy-centered teaching also promotes ethical and social awareness by helping students reflect on the societal impact of their technical decisions and the fairness of software systems (A03, A34). Reflective and experiential learning further enhance motivation and engagement, encouraging students to connect design choices with human outcomes (A03, A34). Some teaching practices focus on bias awareness and inclusion, guiding students to identify accessibility barriers and create equitable technological solutions (A34, A40, A42, A43). Teaching empathy in software education also supports professional growth by developing self-awareness, communication skills, and employability within collaborative and user-focused contexts (A03, A05, A09, A38).

Engineering and software education both show that teaching empathy enriches learning by reinforcing collaboration, ethical reflection, and social responsibility. Engineering programs use empathy to strengthen moral reasoning and complex problem-solving, while software education applies empathy to develop user-centered thinking and responsible design practices. These patterns are summarized in Table \ref{tab:benefits-comparison}.

\begin{table}[t]
\centering
\caption{Comparison of reported learning benefits of teaching empathy in Engineering and Software Engineering Education}
\label{tab:benefits-comparison}
\footnotesize
\begin{tabular}{p{1.6cm} p{2.9cm} p{2.9cm}}
\toprule
\textbf{Benefit Theme} & \textbf{Engineering (General)} & \textbf{Software Engineering} \\
\midrule
Problem Solving & Teaching empathy helps students approach problems with broader context and social awareness & Empathy teaching enhances understanding of users and strengthens design quality \\
\midrule
Collaboration & Learning empathy improves teamwork and communication in group activities & Empathy teaching supports collaboration and communication in design teams \\
\midrule
Ethical Awareness & Teaching empathy increases reflection on ethical responsibilities and social impact & Empathy teaching promotes fairness and responsibility in software development \\
\midrule
Motivation & Empathy activities make learning more meaningful and engaging for students & Empathy-based projects increase motivation and connection to real users \\
\midrule
Bias Awareness & Empathy learning helps students recognize and reduce bias in design decisions & Teaching empathy develops awareness of accessibility and inclusive design \\
\midrule
Professional Growth & Empathy education builds self-awareness, communication, and professional skills & Learning empathy develops reflective and interpersonal competence for teamwork \\
\midrule
Societal Impact & Empathy learning helps students understand engineers’ roles in society & Teaching empathy encourages user-centered and responsible technology creation \\
\bottomrule
\end{tabular}
\end{table}

\subsection{RQ4: Challenges of Teaching and Learning Empathy in Engineering and Software Engineering Education}

In engineering education, several recurring challenges hinder the effective incorporation of empathy into teaching and learning. A common concern across studies is the low prioritization of empathy in relation to technical competence. Engineering programs often emphasize hard skills, while empathy remains undervalued or treated as secondary within curricula (A01, A11, A13, A15, A19, A26, A27, A31, A33, A37). This imbalance reflects a persistent perception that technical mastery defines engineering success, leaving limited room for interpersonal and reflective learning.

Measurement also presents a major barrier. Studies report difficulties in evaluating empathy consistently and distinguishing it from related constructs such as sympathy or moral reasoning (A08, A15). Educators face uncertainty in selecting valid instruments or metrics to assess changes in students’ empathic capacity. Alongside these structural issues, student resistance and limited awareness emerge as significant pedagogical obstacles. Learners often struggle to engage emotionally with design contexts, show reluctance to step outside technical comfort zones, or find classroom simulations unrealistic (A15, A17, A19, A33).

Curricular and resource constraints further limit sustained empathy practice. Short course durations, overloaded programs, and a lack of institutional support make it difficult to embed empathy exercises in meaningful and repeated ways (A15, A17, A32, A36). Cultural and disciplinary barriers also restrict progress. Deeply ingrained professional norms, existing biases, and limited collaboration across social or human-centered disciplines reduce opportunities for transformative empathy learning (A08, A14, A15, A18, A20, A30, A32). Even when empathy activities are introduced, implementation gaps persist, as some classroom practices fail to connect with real-world professional contexts or overlook individual differences among students (A01, A33).

In software engineering education, similar barriers appear but with distinct disciplinary characteristics. A recurring issue involves the undervaluing of empathy in computing programs, where technical expertise is prioritized and human-centered skills receive limited attention (A03, A34, A38, A09). Students often enter courses with low empathy awareness and struggle to shift from analytical reasoning toward human-centered reflection (A02, A05, A09, A38). Measuring empathy also remains difficult, as instructors face challenges defining, quantifying, and observing empathic learning outcomes (A05, A34).

Curricular and resource constraints restrict long-term implementation, with short modules or isolated activities limiting the depth and continuity of empathy development (A02, A03, A05, A34, A38). The technical structure of computing programs often leaves little space for extended experiential learning. Cultural and disciplinary barriers also emerge in multidisciplinary collaborations, where differing vocabularies, expectations, and priorities create friction between technical and design-oriented students (A02, A03). Other difficulties involve the lack of robust pedagogical frameworks for integrating empathy into software education and the limited adoption of such approaches across institutions (A34). Some courses risk superficial engagement, where empathy becomes a procedural exercise rather than a sustained reflective practice (A03, A05, A38).

Both areas reveal that teaching empathy faces systemic and practical challenges. Engineering programs struggle to balance technical rigor with interpersonal development, while software education contends with curricular limitations and cultural resistance within computing disciplines. These barriers demonstrate that empathy teaching requires consistent institutional support, valid assessment strategies, and pedagogical continuity to move from isolated activities to sustained professional competence. The main challenges identified are summarized in Table \ref{tab:challenges-comparison}.

\begin{table}[t]
\centering
\caption{Comparison of reported challenges in teaching and learning empathy in Engineering and Software Engineering Education}
\label{tab:challenges-comparison}
\footnotesize
\begin{tabular}{p{1.6cm} p{2.9cm} p{2.9cm}}
\toprule
\textbf{Challenge Theme} & \textbf{Engineering (General)} & \textbf{Software Engineering} \\
\midrule
Low Prioritization & Empathy undervalued and overshadowed by technical skills in curricula & Empathy receives limited attention compared to technical content in computing courses \\
\midrule
Measurement & Assessing empathy remains complex and lacks consistent instruments & Measuring empathy and engagement is methodologically difficult and often unclear \\
\midrule
Student Resistance & Students struggle to empathize, resist non-technical learning, or find exercises unrealistic & Students show low awareness and difficulty shifting from analytical to empathic reasoning \\
\midrule
Curricular Constraints & Overloaded programs and short courses hinder sustained empathy learning & Limited curricular space and short-term modules restrict long-term empathy development \\
\midrule
Cultural and Disciplinary Barriers & Professional norms, biases, and disciplinary boundaries limit empathy integration & Multidisciplinary tensions and technical focus reduce empathic collaboration \\
\midrule
Implementation Gaps & Classroom activities often disconnected from real practice and may backfire & Few structured frameworks for integrating empathy across software curricula \\
\midrule
Depth of Learning & Time pressure and surface-level tasks reduce reflection and meaningful learning & Empathy sometimes taught as procedural or compliance-based rather than reflective \\
\bottomrule
\end{tabular}
\end{table}

\section{Discussion}
\label{sec:discussion}
This section compares our results with the existing literature and discusses the implications of our findings for research and practice in software engineering education.

\subsection{Comparing Findings with the Literature}
Empathy has long been regarded as an important aspect of professional formation across disciplines, yet its systematic incorporation into engineering education remains uneven. Earlier studies in engineering positioned empathy as a reflective and ethical competency that connects technical reasoning with human and societal contexts \cite{bearman2015learning, van2012based}. This view aligns with long-standing pedagogical traditions in healthcare, nursing, and social sciences, which established empathy as a teachable capacity that supports communication, ethical awareness, and professional responsibility \cite{meek1957experiment, stepien2006educating, jeffrey2016empathy, han2018review, kelley2013teaching}. Engineering education gradually integrated these insights into reflective and ethics-based pedagogies, emphasizing moral reasoning, social awareness, and civic engagement \cite{bearman2015learning, van2012based, gates2023world}. In software engineering, these same principles have been adapted into design-oriented and measurable approaches that embed empathy within concrete development practices. Research on human-centered and design thinking methods describes empathy as a process that bridges user understanding and system functionality through structured activities such as empathy maps, personas, user interviews, and accessibility evaluation \cite{levy2018importance, ferreira2015eliciting, cerqueira2023thematic, clear2024software, gunatilake2023empathy}.

Findings from our review are consistent with these perspectives while refining their disciplinary boundaries. In general engineering education, empathy often appears as a moral and reflective value linked to ethical formation and social responsibility, reinforcing earlier claims that engineers should engage with the human implications of their work \cite{bearman2015learning, van2012based, bearman2015learning, jeffrey2016empathy}. In contrast, software engineering tends to operationalize empathy as a structured pedagogical construct integrated into requirements analysis, user-centered design, and collaborative development \cite{cerqueira2023thematic, clear2024software, devathasan2025empathy}. Both domains share a commitment to social responsibility and inclusive learning but differ in pedagogical focus. The engineering programs identified in this study (i.e., biomedical, civil, chemical, mechanical, and multidisciplinary) emphasize reflective dialogue, ethics discussions, and community engagement \cite{van2012based, thompson1983empathy, bearman2015learning}, whereas software programs emphasize applied methods that link empathy directly to technical outcomes and assessable learning results \cite{ferreira2015eliciting, levy2018importance, cerqueira2024empathy}. This divergence reflects broader disciplinary orientations: engineering frames empathy as part of civic and professional identity, while software education translates it into a measurable design competence aligned with usability, accessibility, and responsible innovation.

These findings contribute to the literature by clarifying how empathy may evolve from a reflective disposition to an operational and pedagogically structured element within technical education. Earlier research emphasized either moral grounding \cite{meek1957experiment, stepien2006educating, jeffrey2016empathy} or design application \cite{levy2018importance, cerqueira2023thematic, clear2024software}, but few examined how these dimensions interact across disciplines. The synthesis presented here indicates that engineering education tends to provide ethical orientation that frames empathy as a professional disposition, while software education embeds that disposition into methods, measurable indicators, and collaborative workflows. Collectively, these perspectives suggest that empathy can function both as a moral compass guiding professional responsibility and as a design instrument informing user-centered and socially responsive software development. This interpretation complements existing frameworks in design thinking and soft skills research \cite{malinen2025soft, matturro2019systematic, lecca2025curious} by showing how empathy contributes to both ethical formation and technical quality in engineering education.

At the same time, our study supports ongoing observations about the challenges of sustaining empathy-based education. Engineering programs continue to prioritize technical mastery, which can leave interpersonal development undervalued \cite{thompson1983empathy, jeffrey2016empathy}. Educators report persistent difficulties in assessing empathic growth and distinguishing empathy from related constructs such as moral reflection or communication skills \cite{stepien2006educating, han2018review}. Similar challenges appear in software education, where curricular structures and limited teaching time often constrain empathy exercises to short-term or isolated activities \cite{cerqueira2023thematic, clear2024software}. Studies also indicate that individuals working with software may struggle to shift from analytical reasoning to perspective-taking, echoing earlier findings about low empathy awareness in technical disciplines \cite{khakurel2020effect, gunatilake2023empathy, devathasan2024deciphering}. Addressing these barriers in software engineering education may therefore require pedagogical frameworks that embed empathy as a continuous and assessable component of technical training rather than as a peripheral addition to design or ethics courses.

The contribution of this study lies in linking two previously separate bodies of work: empathy education in engineering and empathy practice in software development. Prior research emphasized the ethical and social significance of empathy in professional education \cite{meek1957experiment, jeffrey2016empathy, bearman2015learning}, while more recent studies in software engineering have explored its methodological and cognitive potential in user-centered innovation \cite{cerqueira2023thematic, clear2024software, devathasan2025empathy}. By drawing these perspectives together, the analysis points to a gradual convergence in which empathy serves both as an ethical value and as a design capability. The review indicates that software engineering is becoming an active area for operationalizing empathy, as reflected in the emergence of structured frameworks, measurable interventions, and inclusive design practices in recent years \cite{cerqueira2024empathy, devathasan2025empathy, gunatilake2024enablers}. This development suggests a shift from conceptual advocacy toward practice-oriented integration, reinforcing theoretical claims about empathy’s teachability and professional relevance. Beyond documenting this trend, the study offers an empirically grounded view of how ethical reflection and design instrumentation can jointly support empathy as a technical and moral competence. These insights provide a foundation for future curriculum design and research that continue to integrate empathy into software engineering education in ways that are pedagogically consistent and socially meaningful.

\subsection{Implications}
\label{sec:implications}

Our review produced several implications for research and educational practice in software engineering. \newline

\noindent \textbf{Implications for Research}.
Our findings contribute to ongoing discussions in software engineering education by suggesting that empathy is evolving from a peripheral soft skill into a structured pedagogical and methodological construct. Earlier work often described empathy primarily as a design principle, yet limited attention has been given to its role as a measurable educational outcome or as a mechanism for ethical awareness in development processes. The synthesis here indicates that empathy can complement technical competence by supporting collaboration, inclusivity, and user understanding. Future research could examine how empathy-based learning influences problem-solving, bias mitigation, and fairness in software systems, as well as how empathy development interacts with team dynamics and professional identity formation among students. Longitudinal and mixed-methods studies may help trace how empathy develops across educational stages and project contexts, strengthening theoretical foundations for its sustained inclusion within software engineering education. \newline

\noindent \textbf{Implications for Educational Practice}.
Our findings suggest a need to integrate empathy as a continuous and assessable component of software engineering curricula rather than as a short-term or supplementary activity. Empathy learning can be embedded throughout design, requirements, and development courses through structured and iterative exercises. Reflection and feedback mechanisms can reinforce these activities, helping students connect technical decisions with human outcomes. Educators may also consider assessment strategies that combine reflective evaluation with artifact-based measures to make empathy development visible, measurable, and relevant to software quality. Faculty development and institutional support remain important for enabling instructors to design learning environments where empathy is recognized as both a technical and ethical capability central to responsible and inclusive software development.
\section{Conclusions and Future Work}
\label{sec:conclusions}

In this study, we conducted a systematic review to examine how empathy has been incorporated into engineering and software engineering education. Following established guidelines for evidence-based research, we analyzed 43 studies across multiple engineering disciplines to synthesize how empathy is used, taught, and evaluated as a professional and pedagogical construct. The review mapped key patterns in conceptualization, teaching practices, reported benefits, and challenges, suggesting that empathy has gradually evolved from an ethical and reflective ideal toward a teachable competence embedded in technical curricula.

The findings suggest that general engineering programs tend to frame empathy as a moral and social responsibility, while software engineering programs adapt these principles into structured, design-oriented, and assessable pedagogical practices. This development points to an ongoing effort to integrate empathy within user-centered design, collaborative development, and inclusive technology education. The review contributes an empirical foundation that helps to explain how empathy can operate both as a moral compass guiding professional responsibility and as a design instrument informing technical and social outcomes in software engineering education.

Future work will build on these results by investigating how empathy-based interventions can be embedded across the lifecycle of software education. Planned activities include a survey with professors and instructors to understand current practices, barriers, and perceptions related to teaching empathy in software engineering. The survey results will inform the design of an intervention study to evaluate structured empathy-building activities integrated into software engineering courses. These next steps aim to develop frameworks and instructional resources that support educators in assessing empathy systematically and linking it to measurable improvements in student learning and professional formation.

\section*{Data Availability}
The data used in this review is available at \url{https://figshare.com/s/3068ab3422a9e5bf2fc8}. 

\bibliographystyle{ACM-Reference-Format}
\bibliography{bibliography}

\appendix

\end{document}